\documentclass[usenatbib]{article}
\usepackage{times,graphicx,a4wide,natbib}

\begin{document}

\title{A Failure of Serendipity: the Square Kilometre Array will struggle to eavesdrop on Human-like ETI}
\author{D.H. Forgan$^1$ and R.C. Nichol$^2$}
\maketitle

\noindent $^1$Scottish Universities Physics Alliance (SUPA), Institute for Astronomy, University of Edinburgh, Blackford Hill, Edinburgh, EH9 3HJ \\
\noindent $^2$Institute of Cosmology and Gravitation, Dennis Sciama Building, Burnaby Road, Portsmouth, PO1 3FX 

\noindent \textbf{Word Count:} 2,994 \\

\noindent \textbf{Direct Correspondence to:} \\
D.H. Forgan \\ \\
\textbf{Email:} dhf@roe.ac.uk \\
\textbf{Post:} Mr Duncan Forgan \\
Room C18, Institute for Astronomy \\
University of Edinburgh \\
Blackford Hill \\
EH9 3HJ, UK

\newpage

\begin{abstract}

The Square Kilometre Array (SKA) will operate in frequency ranges often used by military radar and other communications technology.  It has been shown that if Extraterrestrial Intelligences (ETIs) communicate using similar technology, then the SKA should be able to detect such transmissions up to distances of  $\sim$ 100 pc ($\sim 300$ light years) from Earth.  However, Mankind has greatly improved its communications technology over the last century, dramatically reducing signal leakage and making the Earth ``radio quiet''.  If ETIs follow the same pattern as the human race, will we be able to detect their signal leakage before they become radio quiet? We investigate this question using Monte Carlo Realisation techniques to simulate the growth and evolution of intelligent life in the Galaxy.  We show that if civilisations are ``human'' in nature (i.e. they are only ``radio loud'' for $\sim 100$ years, and can only detect each other with an SKA-like instrument out to 100 pc, within a maximum communication time of 100 years), then the probability for such civilisations accidentally detecting each other is low ($\sim 10^{-7}$), much lower than if other, dedicated communication techniques are permissible (e.g. optical SETI or neutrino communication).  \\


\end{abstract} 

\newpage

\section{Introduction}

\noindent It is generally believed that communications from an extraterrestrial intelligence will come in two possible forms. The first would be a civilization more advanced than ours that has the technology and power to broadcast signals across the Galaxy, specifically for others to detect (this is a "beacon"). The second would be technologically younger civilizations, like ours, that are just beginning to use advanced communications, which then leak into space for others to eavesdrop on. As outlined by \citet{Penny04}, radio frequencies are believed to be the most natural place to look for both types of these signals and therefore, a majority of past, present and future searches for extraterrestrial intelligence are in the radio (or microwave) range of the electromagnetic spectrum. 

The most ambitious search so far for extraterrestrial beacons was the Phoenix Project, which ran for nearly ten years (from 1995 to 2004), and observed 800 stars (out to 240 light years from earth) with Arecibo, Parkes and the Green Bank radio telescopes (over a frequency range of 1.2 to 3 GHz). Alternatively, the BETA project used a 26m radio telescope to perform an all-sky, narrow-band, microwave search for extraterrestrial beacons in the so-called ``water hole" between 1400-1720 MHz. This range corresponds to a gap in the radio noise spectrum coming from space located between the hydrogen line and the strongest hydroxyl line (water), and it is believed an intelligent race would pick this region of the electromagnetic spectrum to broadcast their beacons. For both experiments, no unusual radio signals were detected and limits were placed on the possible strength of the beacons being broadcast.	

A new era for the search for extraterrestrial Intelligence (or SETI) has begun with the construction of the Allen Telescope Array (ATA), built to continue the search for ``beacons'' across the Galaxy. The first 42 dishes of the ATA have already begun scientific operations, with a further 300 dishes planned. The ATA is targeting $\sim 250,000$ stars (including stars with known exoplanets) in the ``Water Hole'' looking for alien beacons, while also doing a deep blind survey (20 square degrees) towards the Galactic center looking for strong beacons from billions of stars in that direction.

In addition to the dedicated ATA, other radio telescopes are being built across the world, which could also facilitate SETI. For example, the Low-Frequency Array (LOFAR) in Europe has just started exploring the Megahertz regime of the radio spectrum in search of the redshifted 21cm line of neutral hydrogen from the Epoch of Re-ionization. Ironically, LOFAR will encounter significant Radio Frequency Interference (RFI) from modern human activity (radar, mobile phones, radio and TV stations, etc.), and dedicated SETI observations have tended to avoid these radio frequencies because of this problem (working at Gigahertz frequencies instead).

Recently, \citet{Loeb2007} have highlighted the irony of this approach and noted that the increased sensitivity of new radio arrays like LOFAR could eavesdrop on the RFI produced by distant extraterrestrial civilizations. The attraction of this idea is simple: First, the human race is not broadcasting a radio beacon (at Gigahertz frequencies) for other civilization to detect as the power required is prohibitive. Secondly, we are radiating significant radiation from everyday activities in the Megahertz region of the electromagnetic spectrum, so we know of at least one (advanced) civilization that could, in theory, be detected at these radio frequencies (humans). Finally, SETI could easily ``piggy-back'' on the existing operations of LOFAR and other radio arrays. 

In their paper, Loeb and Zaldarriaga focused on the radio emission potentially produced by human-like military radar, which is one of the most powerful sources of radio ``leakage'' into space from Earth. The radar employed by the US Ballistic Missile Defense System (BMDS) can generate isotropic radiation with a total power of 2 billion Watts, or two orders of magnitude higher if beamed. Likewise, over-the-horizon radar, which bounces signals off the ionosphere, can reach similar power output \citep{Tarter2004}. Using such signals as a blueprint for possible extraterrestrial radio emission, Loeb and Zaldarriaga estimated that LOFAR could detect civilizations like ours out to a distance of 50 parsecs with a month of observation (see Figure 1 of their paper). This volume of the Galaxy contains $\sim ~10^5$ stars and several possible rocky exoplanets, e.g., Gliese 581c \citep{Udry2007}.  Beyond LOFAR, plans are already underway for the construction of a Square Kilometer Array (SKA) radio telescope, which Loeb \& Zaldarriaga show could see human-like radio signals to $\sim 200$ pcs (in one month of observation).

However, we should think carefully about what we might expect the SKA to see.  While humans are still leaking radio emission into the Galaxy, the extent of this emission has diminished.  Technological improvements have reduced the transmission power required to broadcast, and the dawn of the digital age has begun to supersede traditional radio entirely.  These events have occurred in just over 100 years, putting us on the path to becoming a ``radio quiet'' civilisation.  If the Biological Copernican Principle is true (i.e. humans are not atypical as intelligent species), then what happens if \emph{all} civilisations rapidly become radio quiet?

In this paper, we expand on the ideas of \citet{Loeb2007} to assess the likelihood of eavesdropping on the radio emission from other advanced civilizations. To achieve this, we use the latest statistical model for likely habitable planets in the Galaxy (see \citealt{mcseti1} for details) combined with an estimation for the likely life-time of radio leakage into space. Together, these provide a probability for being close enough to an advanced civilization, which is broadcasting, for the SKA to detect it.

\section{Numerical Methods}

To study the connectivity of civilisations in the Galaxy, we employ Monte Carlo Realisation Techniques to generate statistical populations of ETIs \citep{mcseti1,mcseti2,entropy}.  The method generates a mock Milky Way of $N$ stars, with properties randomly selected from appropriate statistical distributions - initial mass function \citep{IMF}, star formation history, age-metallicity relation \citep{Rocha_Pinto_SFH}, etc.  Stars are then selected as hosting planetary systems based on their metallicity using empirical relations \citep{Wyatt_Z}.  These planetary systems in turn are generated using statistical distributions to reflect current planet formation theory (a detailed discussion of how this is done is given in \citealt{mcseti2}).  This provides a catalogue of planets, some of which will be potential niches for life.  We then specify a hypothesis for life's formation, so that these niches can be identified.  The hypothesis is essentially a set of selection criteria; planets that pass the selection criteria are seeded with life, which then evolves towards intelligence (modelled by stochastic equations, see \citealt{mcseti1} for more).  The resultant output is a distribution of intelligent civilisations, which arise at various locations and times throughout cosmic history (see Figure 1).  This is referred to as one Monte Carlo Realisation (MCR).  The simulation can then be re-run to produce multiple MCRs, allowing us to average the resulting distributions, and quantify the random uncertainty in the modeling process (though unfortunately the true error in the model selection itself is not accounted for, \citealt{mcseti1}).

Each MCR produces $N_{civ}$ worlds which harbour an intelligent communicating civilisation.  As we store data on all $N_{civ}$ individual civilisations for each MCR, connectivity data can be collected on the number of Intelligent Civilisation Pairs (ICPs) given by $N_{civ}(N_{civ}-1)/2$ in each MCR.  The variables that can then be calculated are:

\begin{enumerate}

\item $dx$, the separation of the two civilisations that compose the ICP,
\item $dt$, the communication interval of the ICP (i.e. the maximum time interval in which both civilisations are extant and communicating),
\item $f$, the contact factor, which identifies the number of “conversations” (pairs of signals) that the ICP can exchange (where $c$ is the speed of light)

\begin{equation} f = \frac{c\, dt}{2\,dx}. \end{equation}
\end{enumerate}  

\noindent To study the effect of our SKA constraint on connectivity, two sets of calculations are required to compare relative trends, as absolute trends are more susceptible to errors in the model selection process.  Both calculations will use the same ensemble of civilisations generated using a relatively optimistic hypothesis of life, merely that life will arise on any planet within the stellar habitable zone (i.e., the zone around each star where the temperature is right for liquid water to be on the surface of any planet).  This ensemble corresponds to the results of the “Baseline” Hypothesis (described in \citealt{mcseti2,entropy}).

\begin{figure}
\begin{center}
\includegraphics[scale=0.7]{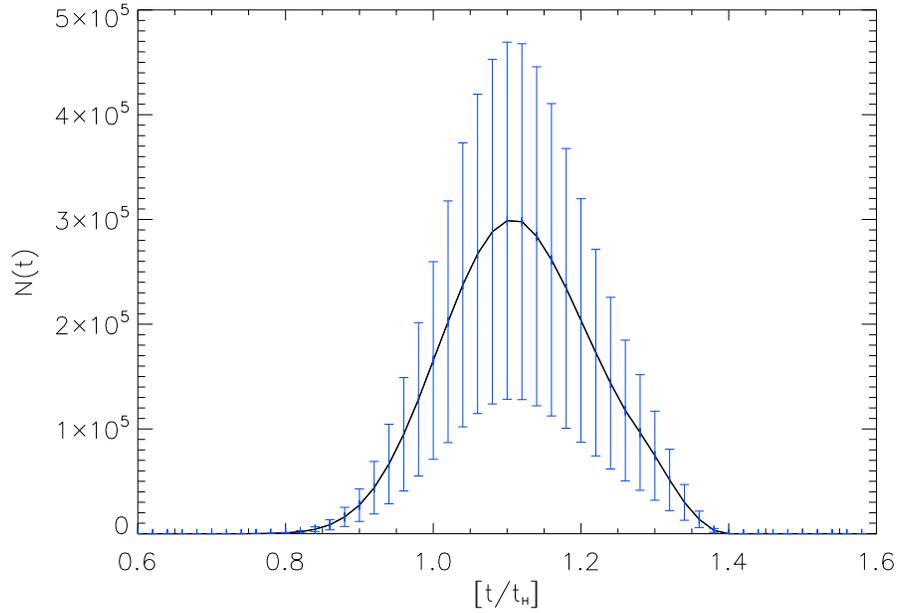}
\caption{The signal history of the Baseline Hypothesis.  The time axis is scaled in units of the Hubble Time (i.e. $t=1 t_h$ indicates the present day).  The black curve represents the mean from 30 individual Monte Carlo Realisations (MCRs) - the blue error bars indicate the standard deviation. \label{fig:signal}}
\end{center}
\end{figure}

A total of 30 MCRs were run for this hypothesis, giving a mean signal number of $N_{civ} \sim 5 \times 10^5$.  This provides an \emph{ab initio} optimistic data set: these simulations have a substantial population of civilisations existing at a similar time in cosmic history (Figure \ref{fig:signal}), offering us the best chance to communicate with ETIs, either by accident (eavesdropping) or using beacons. 

The first calculation of connectivity will use ``unconstrained'' values of $dx$ and $dt$ to calculate the contact factor (i.e. there is no maximum distance over which civilisations can communicate, and the time limit in which civilisations can communicate is the maximum available time, i.e. the communications interval).  These civilisations therefore use a variety of different instruments and techniques to conduct their search at various wavelengths and energy scales.

The second calculations will use ``constrained'' values of $dx$ and $dt$, to emulate the constraints imposed by observing human-like radiation with an SKA-like instrument.  More rigorously, for this ``eavesdropping'' to take place, we require $dx \leq 100$ pc and $dt \leq 100 $ yr. Therefore, civilisations that are separated by more than 100 pc can not be seen (in a month of observation), and we impose a limit of 100 years on the timescale for ``leaking'' radio emission into space. This latter constraint is based upon likely human trends towards more directed (energy efficient) modes of communication (e.g. high-speed internet via optical fibres). We thus assume humans will cease using radio communications in $\sim 100$ years and will no longer be visible in the radio regime of the electromagnetic spectrum.

\section{Results}

\begin{figure}
\begin{center}
\includegraphics[scale=0.7]{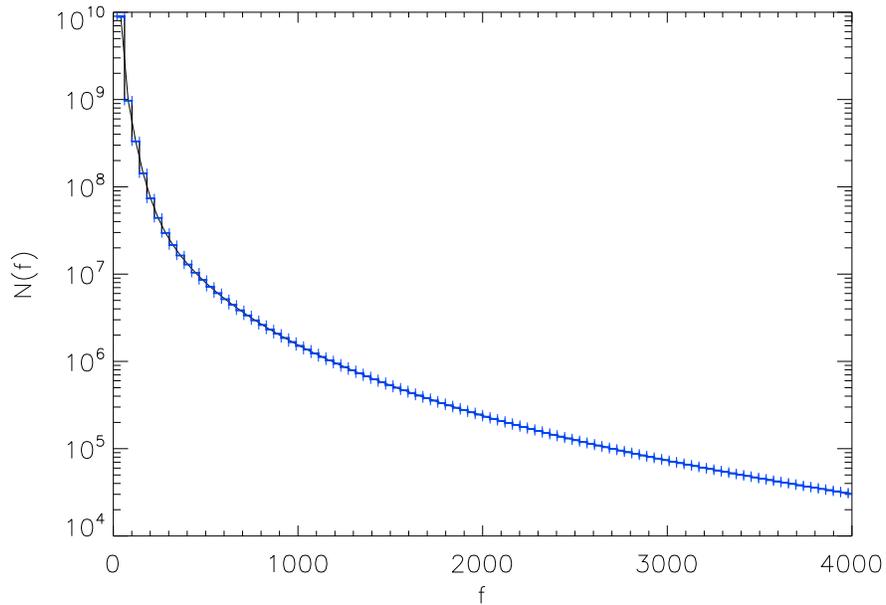}
\caption{Distribution of contact factor for the civilisations in the Baseline Hypothesis, assuming they are radio loud for their entire existence, and there are no constraints on the separation of communicating civilisations (except the constraints imposed by the speed of light and the length of the communication interval).  The black curve represents the mean from 30 individual Monte Carlo Realisations (MCRs) - the blue error bars indicate the standard deviation.\label{fig:baseline}}
\end{center}
\end{figure}

In the unconstrained case, it can be seen that communication is relatively straightforward.  Figure \ref{fig:baseline} shows that a substantial number of civilisation pairs can have a significant number of conversations (represented by a large $f$ value) before the communication interval ends.  While shorter contacts (low $f$) are in general favoured over longer contacts (high $f$), there remains plenty of opportunities for ETI to communicate.

In the constrained case, all communication disappears (i.e. the maximum value of $f=0$).  The combined space-time constraints placed on the civilisations by advancing technology and the spatial limitations of the SKA are extremely strong.  Civilisations are typically separated by distances much greater than 100 parsecs, and are therefore unable to communicate.  Civilisations separated by a distance of 100 pc require a minimum communication interval of

\begin{equation} dt = \frac{2\,dx}{c} = 652\, \mathrm{years} \end{equation}

\noindent in order to establish a dialogue.  Constraining the communications interval to be 100 years long \emph{at its maximum} more or less completely removes the opportunity for communication.   Remember that the communication interval refers to the overlapping time interval in which both civilisations can communicate - it can be as short as a year depending on circumstances and the development of the civilisations that constitute the ICP. 

This analysis does not consider the possibility that if only one message is sent and received (rather than two), ICPs will renew their efforts to continue communications by other methods, despite now being radio quiet.  In our formalism, this corresponds to $f=0.5$ (i.e. instead of a pair of signals being exchanged, only one signal is sent and received).  If this occurs, the communications interval is effectively extended beyond 100 years if $f =0.5$ for $dt \leq 100$ yr.  This extension will only marginally affect the results - most ICPs will still never be able to make the requisite half a conversation to instigate communication.  The fact remains that the probability of an exchange being established is drastically reduced by the constraints of working with SKA-like instruments in an increasingly radio-quiet Galaxy.

\section{Discussion and Conclusions}

We have used Monte Carlo Realisation techniques to study the connectivity of intelligent civilisations when a) the civilisations are “radio loud” or are detectable by other means for their entire existence, with no spatial limits on their detectability, and b) the civilisations are “human-like” in nature, and are limited in  their use of communication tools (e.g., radar) and astronomical instruments, similar in capability to the Square Kilometre Array (SKA): We hypothesis that they become “radio quiet” (and hence undetectable) in 100 years, and can only detect other civilations within 100 pc of each other (using an SKA-like telescope). 

Our results show that civilisations in case a) enjoy significant connectivity, allowing thousands of exchanges of signals traveling at light-speed.  In case b), connectivity decreases to virtually zero, showing that a “SKA eavesdropping experiment” would struggle to detect any human-like ETI in the Galaxy.  

We can justify these results by considering the \emph{civilisation formation rate density} $\dot{n}$, defined as the number of civilisations forming in a given volume in a given time interval.  For the SKA to be able to detect human-like civilisations, then $\dot{n}$ must satisfy:

\begin{equation} \dot{n} \geq 2 (100\, pc)^{-3} (100 yr)^{-1} = 2\times 10^{-8} pc^{-3} yr^{-1} \label{eq:rate}\end{equation}

\noindent If we assume that civilisations form across the entire Galaxy at a constant rate over its lifetime ($\sim 10$ Gyr), then we can calculate the total number of civilisations that the Galaxy must produce:

\begin{equation} N \geq \dot{n} \pi R^2z \Delta t \sim 10^{14} \end{equation}

\noindent Where we have taken canonical values for the Galactic Radius R and height z.  If we attempt to model the civilisation formation rate we obtain from our simulations, then an appropriate model is a Gaussian, with a standard deviation $\sigma_t$, and peak formation time $t_0$ , forming in the Galactic Habitable Zone, an annulus at radius $R=7$ kpc with width $\Delta R =2$kpc and height $z = 300$ pc, (cf Lineweaver et al., 2004): 

\begin{equation} n(t) = \frac{N_0}{V} \frac{1}{\sqrt{2\pi \sigma^2_t}} \exp\left(-\frac{(t-t_0)^2}{2 \sigma^2_t}\right) \end{equation}

\noindent Figure \ref{fig:signal} confirms that $t_0\approx 1.1t_H$ and $\sigma_t \approx 0.3 t_H$, where $t_H=13.7$ Gyr is the current age of the Universe. We normalise the curve using our peak signal $N_{peak} = 3\times 10^5$, giving $N_0=N_{peak}\sqrt{2 \pi \sigma^2_t}$.  This gives a peak formation rate of

\begin{equation} \dot{n} \approx 5 \times 10^{-15} pc^{-3} yr^{-1} \end{equation}

\noindent Modelling in this manner allows us to calculate exactly how populous the Galaxy must be for SKA-based communication.  For a civilisation formation rate density that satisfies equation (\ref{eq:rate}) at the present day, we require that

\begin{equation} N_{peak} \approx 10^{12} \end{equation}

\noindent This is considerably larger than the most positive estimates taken from the Drake Equation.  Coupled with our experimental results, it shows that human-like civilisations will find it difficult to see each other using SKA-like instruments alone.  If we assume that galaxies with a peak formation rate matching that of equation (\ref{eq:rate}) will have communicating civilisations with a probability of 1, then we can estimate that our model predicts a probability of communication of $\sim 10^{-7}$.  These results are based on an assumed observation duration of 1 month.  If we extend this to $\sim 10 $ years, then the SKA's detection radius extends to $\sim 1000 $ pc.  Equation (\ref{eq:rate}) then becomes

\begin{equation} \dot{n} \geq 2 (1000\, pc)^{-3} (100 yr)^{-1} = 2\times 10^{-11} pc^{-3} yr^{-1} \end{equation}

\noindent And the probability of detection is increased to $\sim 10^{-4}$, indicating that for $10^5$ civilisations, a handful of contacts may become possible with this detection radius.  However, 10 years of constant observation with the SKA is unlikely, given the plethora of demands for SKA time from many astronomical fields.

While the SKA remains an important instrument for SETI researchers, its abilities are limited to detecting civilisations that are at a “less advanced” state than Mankind, i.e. they must develop radio technology that remains radio-loud for a significant period of time.  This strengthens the argument for a multi-wavelength approach to SETI, as radio-quiet civilisations may be optically loud, or detectable at some other energy scale (cf \citealt{LearnedJ.G.1994,SilagadzeZ.K.2008}). 

Our calculations suggest that accidental communications, through the eavesdropping of radiation, is highly unlikely and therefore would require civilizations to actively beam signals to us to maximize our chances of detecting their existence. This requires a lot of energy and thus demands civilizations much more advanced than humans, and also better placed to manage the ``costs'' of such work \citep{empire}. The beam would most likely be collimated to boost its strength, hence the civilisation must find a target for the beam.  Therefore, they would also need to know we are here, demanding more sophisticated astronomical instruments, e.g. radio telescopes far in excess of the SKA, or optical instruments that can characterise habitable planets beyond even the capabilities of the putative DARWIN TPF flotilla \citep{Ollivier2008}, or the European Extremely Large Telescope \citep{Kaltenegger2010}.

The conclusions of this paper are not new, and it has been appreciated for some time that the discovery of radiation from civilisations at our
present-day level and with short timescales would be unlikely. Our work simply updates these arguments using the latest predictions for the possible distribution of planets harboring life in our Galaxy, and the probable sensitivities of the newest radio arrays. Such a conclusion could also be seen as being rather pessimistic for SETI but like others, we would like to stress that we are guaranteed to find nothing if we give up looking. What these results stress strongly is that SETI must be a multi-wavelength endeavour, conducted with broader horizons \citep{Davies2010} and a better understanding of our own limitations.

\section{Acknowledgements}

\noindent This work has made use of the resources provided by the Edinburgh Compute and Data Facility (ECDF, http://www.ecdf.ed.ac.uk/). The ECDF is partially supported by the eDIKT initiative (http://www.edikt.org.uk).  The authors would like to thank Alan Penny for his helpful comments which significantly improved this paper.

\bibliographystyle{mn2e} 
\bibliography{IJAduncanforgan}

\end{document}